\documentclass[final,5p,times,twocolumn,compress,colorlinks = true,linkcolor = blue,urlcolor = blue,citecolor = blue,anchorcolor = blue]{elsarticle}

\usepackage{amssymb}
\usepackage{amsmath}
\usepackage{graphicx}
\usepackage{float}
\usepackage{dcolumn}
\usepackage{multirow}
\usepackage[utf8]{inputenc}
\usepackage[T1]{fontenc}
\usepackage{enumerate}
\usepackage{times}
\usepackage{microtype}
\usepackage{orcidlink}
\usepackage{stfloats}
\usepackage{array}
\usepackage{braket}

\usepackage{ulem}
\usepackage{xcolor}

\usepackage{todonotes}
\journal{Physics Letters B}

\begin{document}

\begin{frontmatter}


\title{Constraining neutron star properties through parity-violating electron scattering experiments and relativistic point coupling interactions}

\author[first,second]{P.S. Koliogiannis\corref{cor1}\orcidlink{0000-0001-9326-7481}}
\ead{pkoliogi@phy.hr}
\author[third] {E. Y{\"{u}}ksel}
\ead{e.yuksel@surrey.ac.uk}
\author[first]{N. Paar}
\ead{npaar@phy.hr}

\cortext[cor1]{P.S. Koliogiannis}

\affiliation[first]{organization={Department of Physics, Faculty of Science, University of Zagreb, Bijeni\v cka cesta 32},%Department and Organization
            %addressline={}, 
            city={Zagreb},
            postcode={10000}, 
            %state={},
            country={Croatia}}
            
\affiliation[second]{organization={Department of Theoretical Physics, Aristotle University of Thessaloniki},%Department and Organization
            %addressline={}, 
            city={Thessaloniki},
            postcode={54124}, 
            %state={},
            country={Greece}}
\affiliation[third]{organization={School of Mathematics and Physics, University of Surrey},
            city={Guildford, Surrey},
            postcode={GU2 7XH}, 
            %state={},
            country={United Kingdom}}

\begin{abstract}
Parity-violating electron scattering experiments on $\rm ^{48}Ca$ (CREX) and $\rm ^{208}Pb$ (PREX-2) offer valuable insight into the isovector properties of finite nuclei, providing constraints for the density dependence of the nuclear equation of state, which is crucial for understanding astrophysical phenomena. In this work, we establish functional dependencies between the properties of finite nuclei—such as weak charge form factors and neutron skin thickness—and the bulk properties of neutron stars, including tidal deformability from binary neutron star mergers and neutron star radii. The dependencies are formulated by introducing a family of $\beta$-equilibrated equations of state based on relativistic energy density functionals with point coupling interactions. The charge minus the weak form factors derived from CREX and PREX-2 measurements, combined with the observational constraints on tidal deformability from the GW170817 event, are used to constrain the symmetry energy and neutron star radii. Notably, the energy density expanded up to the fourth order in symmetry energy yields larger radii compared to calculations limited to the second order term. However, the results reveal a discrepancy between the constraints provided by the CREX and PREX-2 experiments. For a more quantitative assessment, higher precision parity-violating electron scattering data and neutron star observations are required.
\end{abstract}


\end{frontmatter}

\section{Introduction}
The nuclear matter equation of state (EoS), which governs the properties of neutron stars, is subject to uncertainties that arise mainly from the density dependence of the nuclear symmetry energy, especially at higher densities. The nuclear symmetry energy sets a high sensitivity to the EoS through the isovector channel of nuclear interaction, which plays a key role in understanding asymmetric nuclear matter. In particular, the isovector component of the interaction affects both the macroscopic properties of neutron stars and those of finite nuclei, serving as a bridge between nuclear physics and nuclear astrophysics. Therefore, terrestrial experiments on the structure of finite nuclei provide significant insights into the structure of neutron stars~\cite{STEINER2005325,universe6080119,particles6010003}.

To date, numerous studies have been conducted on finite nuclei to extract possible correlations between nuclear properties, nuclear symmetry energy, and its density dependence [see Refs. \cite{BALDO2016203,ROCAMAZA201896} and references therein]. Particular importance is given to the values of the symmetry energy and its slope around the saturation density. Relevant isovector nuclear quantities include, e.g., neutron skin thickness~\cite{ROCAMAZA201896,PhysRevC.76.051603,PhysRevLett.102.122502,PhysRevC.81.051303,PhysRevC.86.015803,PhysRevC.93.064303}, dipole polarizability \cite{ROCAMAZA201896,PhysRevC.81.051303,PhysRevC.88.024316,PhysRevC.92.064304,PhysRevC.90.064317}, and weak form factor \cite{PhysRevC.88.034325,Erler_2015}. Recently, of particular interest are parity-violating electron scattering experiments in $\rm ^{208}Pb$~\cite{PhysRevLett.126.172502} and $\rm ^{48}Ca$~\cite{PhysRevLett.129.042501}, namely PREX-2 and CREX, respectively. Specifically, the measurement of parity-violating asymmetry allows the extraction of weak form factors and the neutron skin thickness---both quantities strongly linked to the density-dependence of the symmetry energy \cite{PhysRevLett.102.122502,PhysRevC.88.034325}.

Microscopic and macroscopic properties of neutron stars also exhibit a significant dependence on the nuclear symmetry energy, including mainly the composition of matter, neutrino processes relevant to the rapid cooling of neutron stars, reaction rates involved in the astrophysical r-process, as well as the crust's thickness, mass, and radius~\cite{particles6010003}. In addition, astrophysical phenomena, such as the GW170817 event~\cite{Abbott-2019}, describing a binary neutron star merger, probe the low-density behavior of the EoS through measurements of the tidal deformability, which is directly related to the density dependence of nuclear symmetry energy.

In recent years, extensive research has focused on establishing a connection between the PREX-2 and CREX experiments and neutron stars. In Refs.~\cite{PhysRevResearch.4.L022054,PhysRevLett.127.192701,PhysRevC.104.065804,PhysRevLett.126.172503}, the neutron skin thickness value extracted from the PREX-2 measurement has been combined with astrophysical observations to constrain the symmetry energy and its slope. Furthermore, in Refs.~\cite{Biswas_2021,PhysRevC.107.015803,PhysRevC.109.055805} the implications of PREX-2 on the neutron star's bulk properties have been investigated. Additional studies have incorporated both PREX-2 and CREX experiments to assess their implication on nuclear matter as well as neutron star matter, as detailed in Refs.~\cite{PhysRevC.107.055801,PhysRevC.108.024317,YUKSEL2023137622,MIYATSU2023138013,LILIANI2024122812}.

In this letter, we present, for the first time, a family of $\beta$-equilibrated EoSs based on relativistic energy density functionals (EDFs) with density-dependent point-coupling interactions (DD-PC), recently developed in Ref.~\cite{Yuksel2021}, and employ three other EoSs that are consistent with the CREX and PREX-2 experimental data~\cite{YUKSEL2023137622}. EDFs provide a robust framework for investigating the properties of both finite nuclei and neutron stars, incorporating the behavior of nuclear matter across various densities. By capturing the essential aspects of the symmetry energy, EDFs provide valuable insights into nuclear properties, both in terrestrial experiments and astrophysical phenomena. In particular, we aim to connect experimental probes of nuclear matter introduced by the recent parity-violating electron scattering experiments (PREX-2~\cite{PhysRevLett.126.172502} and CREX~\cite{PhysRevLett.129.042501}), 
namely the weak form factors, as well as the corresponding extracted values for neutron skin thickness, to astrophysical observables of neutron stars in the multimessenger era, such as the gravitational mass-radius and dimensionless tidal deformability at $\rm 1.4~M_{\odot}$~\cite{Abbott-2019}. The aforementioned procedure allows us to derive strong constraints originating from the EDF framework with DD-PC interactions on the bulk properties of neutron stars, as well as the symmetry energy, which is explicitly linked to the EoS of dense nuclear matter.

\section{Theoretical Framework}
The research in this work is performed within the relativistic EDF framework, where the nuclear ground-state density and energy are determined through the self-consistent solution of relativistic single-nucleon Kohn-Sham equations~\cite{PhysRev.140.A1133}. This framework is based on a Lagrangian density formulation that incorporates the isoscalar-scalar (S), isoscalar-vector (V) and isovector-vector (TV) four-fermion interactions~\cite{PhysRevC.78.034318,NIKSIC20141808},
\begin{flalign}
\label{Lagrangian}
& \mathcal{L} = \bar{\psi} (i\gamma \cdot \partial -m)\psi
     - \frac{1}{2}\alpha_S(\hat{\rho})(\bar{\psi}\psi)(\bar{\psi}\psi) \nonumber \\
     & \quad-  \frac{1}{2}\alpha_V(\hat{\rho})(\bar{\psi}\gamma^\mu\psi)(\bar{\psi}\gamma_\mu\psi) 
     - \frac{1}{2}\alpha_{TV}(\hat{\rho})(\bar{\psi}\vec{\tau}\gamma^\mu\psi)
                                                                 (\bar{\psi}\vec{\tau}\gamma_\mu\psi) \nonumber \\
    & \quad -\frac{1}{2} \delta_S (\partial_\nu \bar{\psi}\psi)  (\partial^\nu \bar{\psi}\psi) 
         -e\bar{\psi}\gamma \cdot A \frac{(1-\tau_3)}{2}\psi\;. &
\end{flalign}
The point-coupling Lagrangian includes free-nucleon term, point-coupling interaction terms, the coupling of protons to the electromagnetic field, and a derivative term, which accounts for the leading effects of finite-range interactions, required for a quantitative description of nuclear density distribution. The couplings of the interaction terms $\alpha_S$, $\alpha_V$, and $\alpha_{TV}$ are functionals of the vector density $\rho_v=\sqrt{j_{\mu}j^{\mu}}$ with the nucleon four-current $j^{\mu}=\bar{\psi}\gamma_\mu\psi$, to account for medium effects. 
The functional form of the couplings is given by~\cite{PhysRevC.78.034318, NIKSIC20141808},
\begin{flalign}
& \alpha_i(\rho) = a_i + (b_i+c_i x)e^{-d_ix}\quad\quad (i\equiv S, V, TV)\;, &
\label{ansatz}
\end{flalign}
with $x=\rho/\rho_{0}$, and $\rho_{0}$ denoting the nucleon density at saturation in symmetric nuclear matter. Overall, the point-coupling model used in this work has ten parameters: $a_{S}$, $b_{S}$, $c_{S}$, $d_{S}$, $a_{V}$, $b_{V}$, $d_{V}$
$b_{TV}$, $d_{TV}$ and $\delta_S$. The model parameters are constrained through the relativistic Hartree-Bogoliubov (RHB) model to reproduce the properties of open-shell nuclei~\cite{Yuksel2021,PhysRevC.78.034318,NIKSIC20141808}. In translationally invariant infinite nuclear matter, all terms involving the derivative couplings drop out, and the spatial components of the four-current vanish in the Lagrangian \cite{FINELLI2004449}.

For this study, we employ a recently developed family of eight relativistic DD-PC functionals~\cite{Yuksel2021}. These functionals are optimized to consistently describe nuclear ground-state properties across a range of nuclei and systematically vary the symmetry energy at saturation density, $J$, with values ranging from 29 to 36 MeV. Each functional in the DD-PC family is derived through a $\chi^2$ minimization procedure, providing a quantitative description of nuclear binding energies and radii, while also allowing for variation in the symmetry energy parameter $J$ to assess the sensitivity of various nuclear properties on the symmetry energy. This family has been successfully applied in various studies of nuclear ground-state and excitation properties, including dipole polarizability and magnetic dipole transitions~\cite{Yuksel2021, PhysRevC.108.054305}. We denote the eight functionals of this family as DD-PC-J29,...,36. Their model parameters are tabulated in Ref.~\cite{Yuksel2021}.
In addition, we use two functionals, DD-PC-CREX and DD-PC-PREX, which are constrained by the experimental results from the weak form factors $F_W$, measured in the CREX ($^{48}$Ca)~\cite{PhysRevLett.129.042501} and PREX-2 ($^{208}$Pb)~\cite{PhysRevLett.126.172502} experiments, respectively. These form factors are evaluated at the
momentum transfer from the experiment, i.e.,
$\langle Q^{2}\rangle =0.0297~{\rm GeV^{2}}$ from CREX and $\langle Q^{2}\rangle =0.00616~{\rm GeV^{2}}$ from PREX-2~\cite{PhysRevLett.126.172502,PhysRevLett.129.042501}.
We also include the DD-PC-REX functional, which is jointly constrained by weak charge data for both $^{48}$Ca and $^{208}$Pb, providing further information on nuclear matter properties. Finally, the DD-PCX functional \cite{Yuksel2019}, optimized using both isovector dipole and isoscalar monopole excitation properties of ${}^{208}$Pb alongside nuclear ground-state data, offers an independent constraint on the symmetry energy relevant to this study. Overall, this diverse set of functionals enables a comprehensive exploration of nuclear matter properties across different symmetry energies. 

Moving on to the structure of the EoS for neutron star matter, we require that the matter be in a charge-neutral state composed of neutrons, protons, electrons, and muons in $\beta$-equilibrium~\cite{Shapiro-1983}. The calculation of the energy density and pressure for the baryon components of the nuclear matter EoS is implemented as
\begin{flalign}
    & \mathcal{E}_{b}(\rho,\delta) = \rho~E_{b}(\rho,\delta), \\
    & P_{b}(\rho,\delta) = \rho^{2}\frac{\partial E_{b}(\rho,\delta)}{\partial \rho}, &
    \label{eq:energy_density}
\end{flalign}
where $E_{b}(\rho,\delta)$ is the energy per baryon, expressed in terms up to the fourth order in symmetry energy
\begin{flalign}
    & E_{b}(\rho,\delta) = E_{b}(\rho,0) + \delta^{2}S_{2}(\rho) + \delta^{4}S_{4}(\rho), &
    \label{eq:energy_per_baryon}
\end{flalign}
$E_{b}(\rho,0)$ is the energy per baryon of symmetric nuclear matter, $\rho$ denotes the density, and $\delta=(\rho_{n}-\rho_{p})/(\rho_{n}+\rho_{p})$ is the neutron-proton asymmetry parameter. The terms $S_{k}(\rho)$, with $k=2,4$, represent the second and fourth order of the symmetry energy,
\begin{flalign}
    & S_{k}(\rho) = \frac{1}{k!}\frac{\partial^{k} E_{b}(\rho,\delta)}{\partial \delta^{k}}\Bigg\vert_{\delta=0}. &
    \label{eq:symmetry_energy}
\end{flalign}
The symmetry energy $S_{2}(\rho)$ can typically be expanded around the saturation density
\begin{flalign}
    & S_{2}(\rho) = J + Lu + \mathcal{O}(u^{2}), &
\end{flalign}
where $u=(\rho-\rho_{0})/3\rho_{0}$, $J\equiv S_{2}(\rho_{0})$, and $L\equiv 3\rho_{0}\frac{\partial S_{2}(\rho)}{\partial \rho}\vert_{\rho=\rho_{0}}$ represent the symmetry energy and its slope at saturation density $\rho_{0}$, respectively (see Ref.~\cite{BALDO2016203}). However, in this study, the symmetry energy terms are calculated following Eq.~\eqref{eq:symmetry_energy}. The contribution to both the energy density and pressure due to leptons is given by the well-known formula of the relativistic Fermi gas~\cite{Shapiro-1983}.
The aforementioned equations, together with the $\beta$-equilibrium conditions~\cite{Shapiro-1983}, are employed to construct (a) a family of EoSs with symmetry energies ranging from $29$ to $36~{\rm MeV}$ based on the family of DD-PC EDFs, (b) EoSs calibrated to parameters from the CREX and PREX-2 experiments, as well as a combination of both, and (c) an EoS adjusted to meet constraints derived from dipole polarizability measurements.

The mechanical equilibrium of neutron star matter is governed by the well-established Tolman–Oppenheimer–Volkoff (TOV) equations, which are solved self-consistently with the EoS of the fluid interior~\cite{Shapiro-1983,Glendenning-2000}. The TOV equations are expressed as follows:
\begin{flalign}
	& \frac{dP(r)}{dr}=-\frac{G\mathcal{E}(r) M(r)}{c^{2}r^2}\left(1+\frac{P(r)}{\mathcal{E}(r)}\right)\left(1+\frac{4\pi P(r) r^3}{M(r)c^2}\right) \\\nonumber & \quad\quad\quad\quad \left(1-\frac{2GM(r)}{c^2r}\right)^{-1}, \\
    & \frac{dM(r)}{dr}=\frac{4\pi r^2}{c^{2}}\mathcal{E}(r), &
\end{flalign}
providing fundamental properties of neutron stars.

In addition, gravitational wave detectors have accurately measured the effective tidal deformability, $\tilde{\Lambda}$, which is expressed as~\cite{Abbott-2019}
\begin{flalign}
    & \tilde{\Lambda}=\frac{16}{13}\frac{(12q+1)\Lambda_1+(12+q)q^4\Lambda_2}{(1+q)^5}, &
    \label{eftidal}
\end{flalign}
where $q=m_2/m_1\leq 1$ is the binary mass ratio, and $\Lambda_i$ is the dimensionless tidal deformability for each star~\cite{Abbott-2019}
\begin{flalign}
    & \Lambda_i=\frac{2}{3}k_2\left(\frac{R_i c^2}{M_i G}  \right)^5\equiv\frac{2}{3}k_2 \beta_i^{-5}  , \quad i=1,2, &
    \label{dimtidal}
\end{flalign}
with $k_{2}$ denoting the tidal Love number, and $\beta=GM/Rc^{2}$ representing the compactness of the star. $R_{i}$ and $M_{i}$ represent the radii and masses of the components of the binary system, respectively. It needs to be noted that the dimensionless tidal deformability is notably sensitive to the neutron star radius, which is directly influenced by the low-density behavior of the EoS. This behavior reflects the interplay between the microscopic properties of finite nuclei and the macroscopic characteristics of neutron stars. For modeling the crust region of the EoSs, the inner crust is described using the SLy EoS~\cite{Douchin_2001}, which applies in the density range $1.99\times10^{-4}~{\rm fm}^{-3} \leq \rho < \rho_{\rm tr}$. The outer crust follows the EoS established by Baym, Pethick, and Sutherland~\cite{Baym-71}, spanning $6.93\times 10^{-13}~{\rm fm}^{-3} \leq \rho < 1.99\times 10^{-4}~{\rm fm}^{-3}$, while at even lower densities, the Feynman-Metropolis-Teller EoS~\cite{PhysRev.75.1561} is employed for $4.73\times 10^{-15}~{\rm fm}^{-3} \leq \rho < 6.93\times 10^{-13}~{\rm fm}^{-3}$. This framework ensures a consistent and comprehensive description of neutron star structure across different density regimes. The crust-core transition density, $\rho_{\rm tr}$, is determined using the thermodynamical method~\cite{PhysRevC.85.024302,PhysRevC.90.011304}, providing a physically motivated criterion for delineating the transition to the homogeneous core.

\section{Results and Discussion}
\begin{figure}[t!]
\includegraphics[width=\columnwidth]{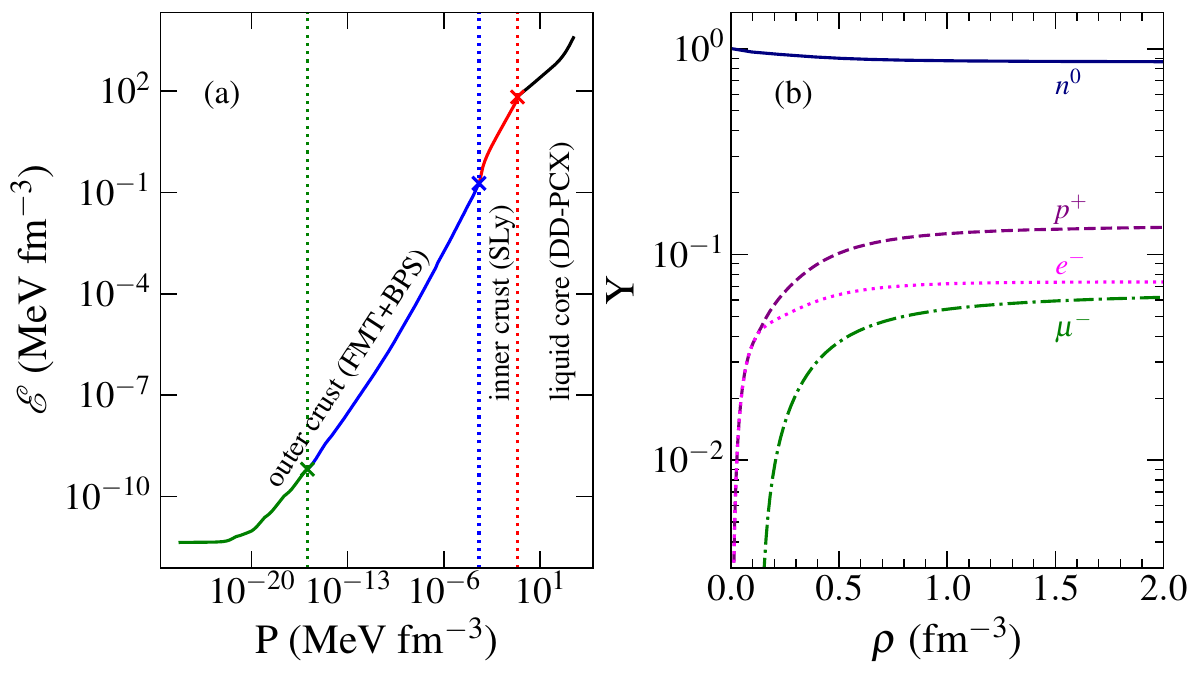}
\caption{(a) Energy density as a function of pressure, highlighting the separate branches corresponding to the outer crust, inner crust, and liquid core, and (b) particle fractions as functions of density in the liquid core. The representative EoS is the DD-PCX.}
\label{fig:eos_branches}
\end{figure}

To visualize the EoSs and the composition of $\beta$-equilibrated neutron star matter, Fig.~\ref{fig:eos_branches} presents (a) the energy density as a function of pressure, distinguishing the separate branches corresponding to the outer crust, inner crust, and liquid core, and (b) the particle fractions as functions of density in the liquid core for the DD-PCX EoS. All considered EoSs exhibit similar behavior.

\begin{figure}[t!]
\includegraphics[width=\columnwidth]{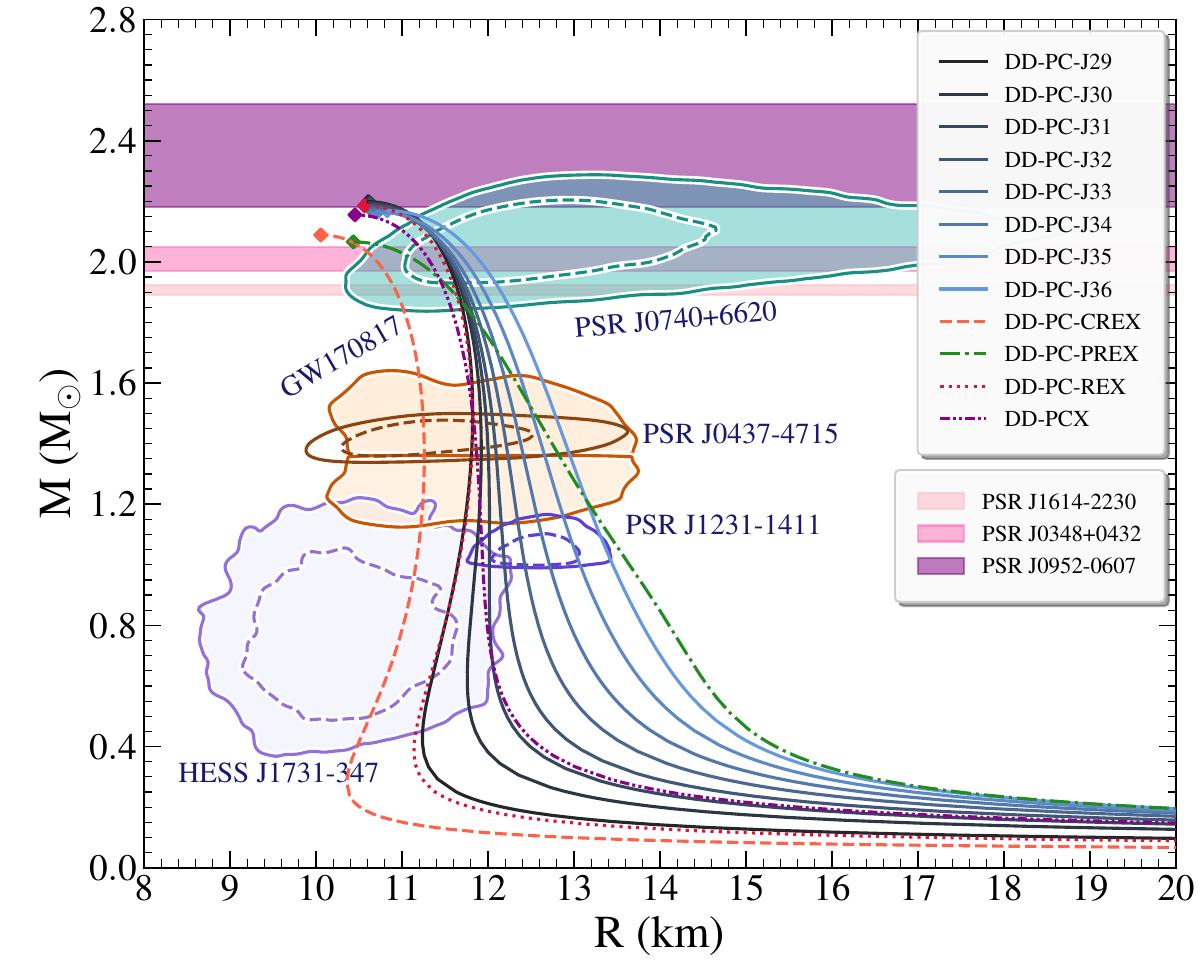}
\caption{Gravitational mass as a function of the radius for the DD-PC EoSs. The solid lines denote the family of DD-PC EoSs spanning $\rm J=29,30,...,36~MeV$, while the dashed, dash-dotted, dotted, and dash-dot-dotted lines correspond to the DD-PC-CREX, DD-PC-PREX, DD-PC-REX, and DD-PCX EoSs, respectively. The shaded contours from bottom to top represent the HESS J1731-347 remnant~\cite{Doroshenko-2022}, the PSR J1231-1411~\cite{Salmi_2024}, the PSR J0437-4715~\cite{Choudhury_2024}, the GW170817 event~\cite{Abbott-2019}, and the PSR J0740+6620~\cite{Fonseca_2021,Dittmann_2024}, while the horizontal shaded regions denote the PSR J1614-2230~\cite{Arzoumanian-2018}, PSR J0348+0432~\cite{Antoniadis-2013}, and PSR J0952-0607~\cite{Romani-2022} pulsar observations with possible maximum neutron star mass.}
\label{fig:mr_ddpc}
\end{figure}

\begin{figure}
\includegraphics[width=\columnwidth]{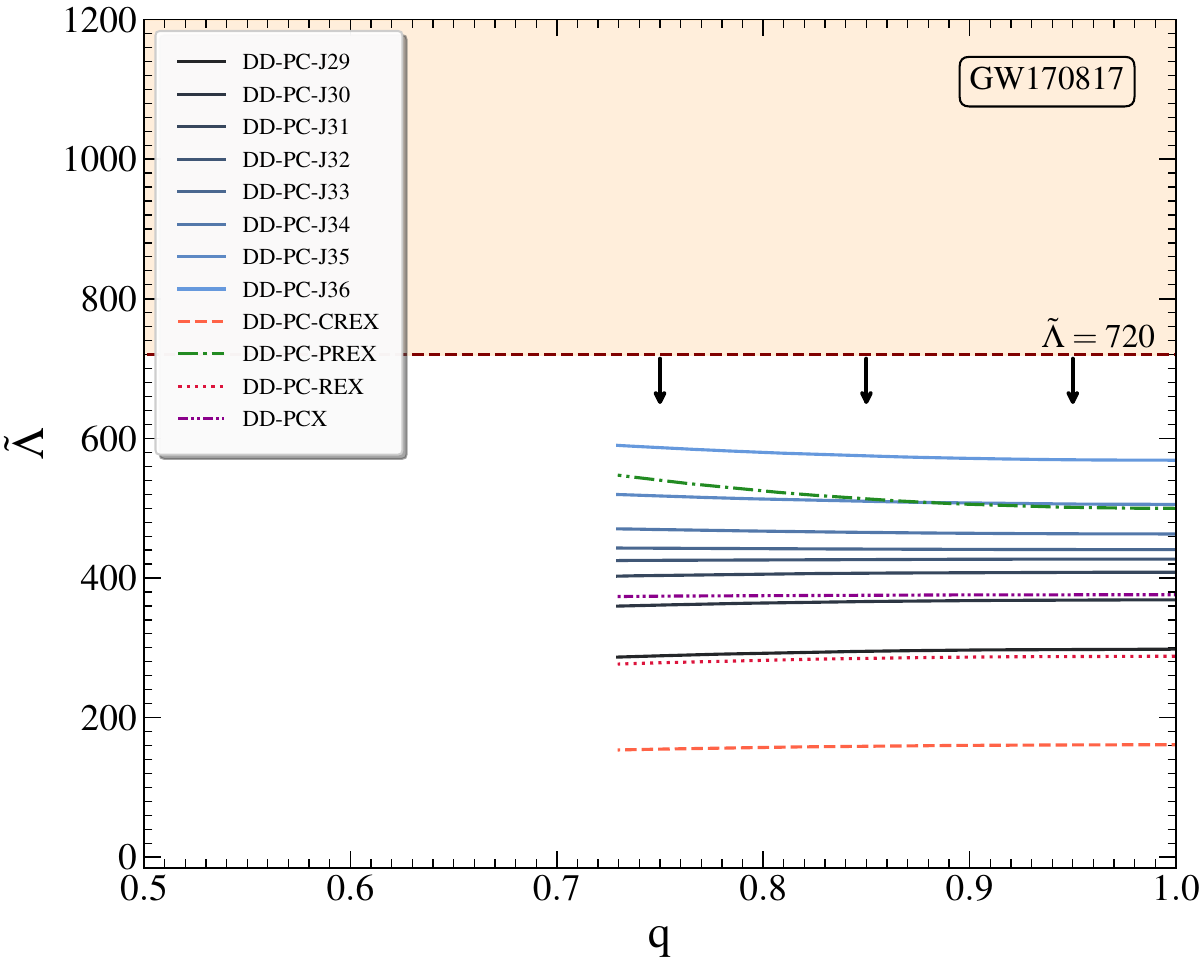}
\caption{Effective tidal deformability as a function of the binary mass ratio for the DD-PC EoSs. The solid lines denote the family of DD-PC EoSs spanning $\rm J=29,30,...,36~MeV$, while the dashed, dash-dotted, dotted, and dash-dot-dotted lines correspond to the DD-PC-CREX, DD-PC-PREX, DD-PC-REX, and DD-PCX EoSs, respectively. The shaded region represents the excluded values provided by LIGO for the GW170817 event~\cite{Abbott-2019}.}
\label{fig:eff_tidal_ddpc}
\end{figure}

Fig.~\ref{fig:mr_ddpc} illustrates the neutron star gravitational mass as a function of the corresponding radius for the DD-PC EoSs. For comparison, mass-radius constraints derived from observational data are illustrated as shaded contours~\cite{Abbott-2019,Doroshenko-2022,Salmi_2024,Choudhury_2024,Fonseca_2021,Dittmann_2024}, and maximum neutron star mass constraints based on pulsar observations denoted as horizontal shaded regions~\cite{Arzoumanian-2018,Antoniadis-2013,Romani-2022}. 
It is noteworthy that all EoSs under consideration conform to the established constraints on the maximum neutron star mass~\cite{Fonseca_2021,Arzoumanian-2018,Antoniadis-2013,Romani-2022}, as well as the mass-radius limits imposed by the observation of the GW170817 event~\cite{Abbott-2019}, and the PSR J0437-4715~\cite{Choudhury_2024}. Additionally, the majority of EoSs fulfill the derived limits from PSR J1231-1411~\cite{Salmi_2024}. Furthermore, a significant subset of these EoSs with lower symmetry energy values $(J=29-32~{\rm MeV})$, successfully describe the characteristics of the ultra-light compact object HESS J1731-347~\cite{Doroshenko-2022}, while simultaneously satisfying the mass-radius constraints from GW170817 and PSR J0740+6620. The GW170817 constraints are further depicted in Fig.~\ref{fig:eff_tidal_ddpc}, given in terms of effective tidal deformability $\tilde{\Lambda}$ from Eq.~\eqref{eftidal}, shown as a function of the binary mass ratio, where all the EoSs also reside within the permitted region defined by the GW170817 event, $\tilde{\Lambda}\leq 720 $.

Focusing on the microscopic properties of finite nuclei constrained by parity-violating electron scattering experiments, as well as the macroscopic characteristics of neutron stars, both having their foundation in the same EDFs, we investigate the potential relationships between these quantities and their implications for deriving neutron star constraints.

\begin{figure*}[t!]
\centering
\includegraphics[width=0.85\textwidth]{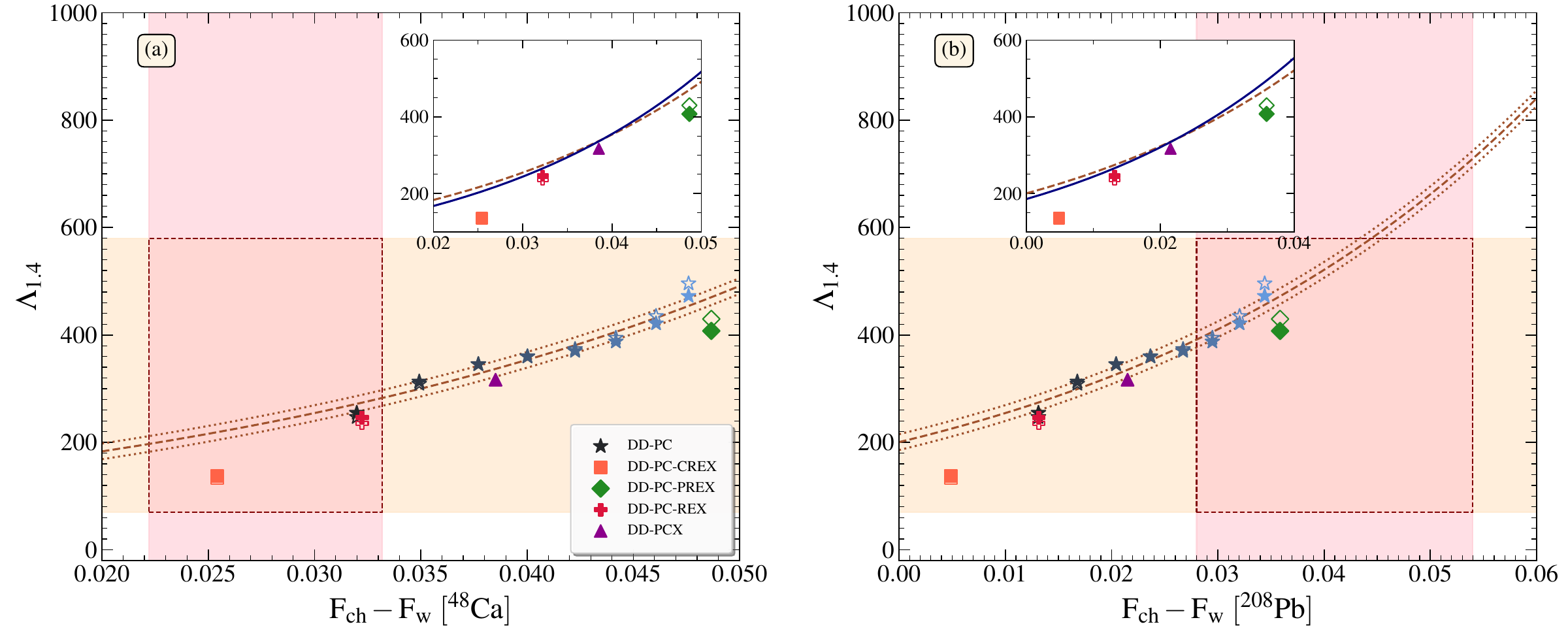}
~
\includegraphics[width=0.85\textwidth]{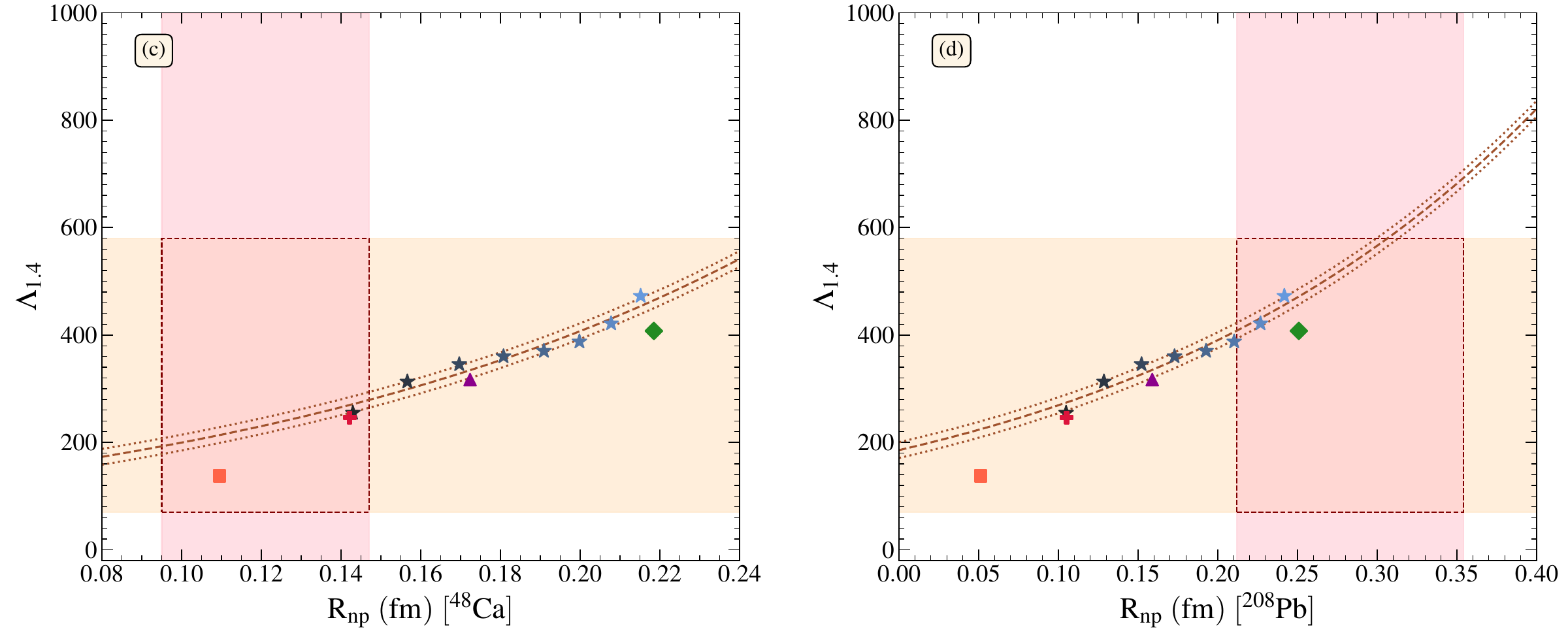}
\caption{Dimensionless tidal deformability at $\rm 1.4~M_{\odot}$ as a function of (a,b) the difference between charge and weak form factors and (c,d) the neutron skin thickness for $\rm ^{48}Ca$ (left panel) and $\rm ^{208}Pb$ (right panel) nuclei. 
The horizontal shaded region represents the constraints from GW170817 event~\cite{Abbott-2019}, while the vertical shaded regions correspond to the experimental constraints extracted from (a,c) CREX and (b,d) PREX-2. The symbols denote different EoSs: DD-PC family of EoSs (stars, where lighter color denotes higher values of $J$), DD-PC-CREX (square), DD-PC-PREX (diamond), DD-PC-REX (plus sign), and DD-PCX (triangle). Open symbols represent the approach considering terms up to $S_{2}(\rho)$ (in some cases the up to $S_{2}(\rho)$ and up to $S_{4}(\rho)$ results overlap). The dashed line represents the fit corresponding to the DD-PC family of EoSs, while the dotted lines denote the standard deviation. The inset plots illustrate the differences between expansions up to the second- and fourth-order terms of the symmetry energy, where the dashed line represents the fit corresponding to the approach considering terms up to $S_{4}(\rho)$, and the solid line the approach considering terms up to $S_{2}(\rho)$.}
\label{fig:L_ddpc}
\end{figure*}

\begin{figure*}[t!]
\centering
\includegraphics[width=0.85\textwidth]{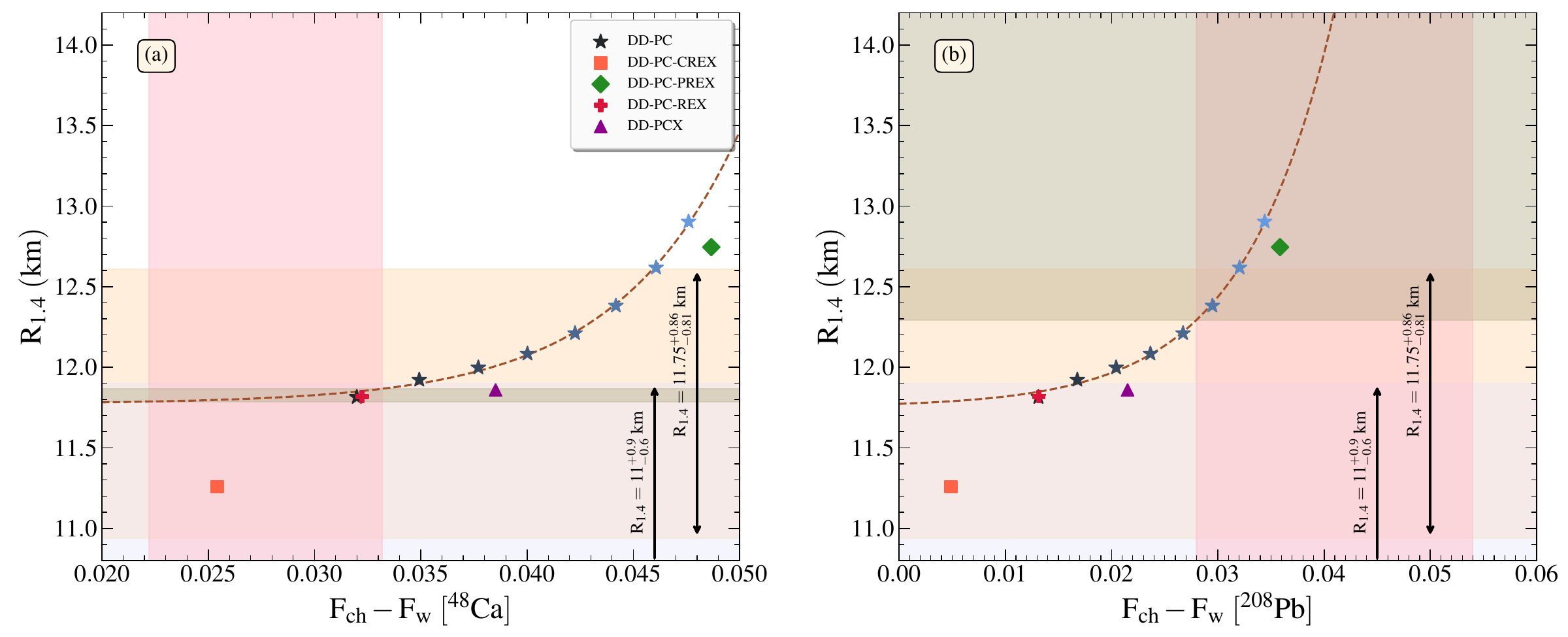}
~
\includegraphics[width=0.85\textwidth]{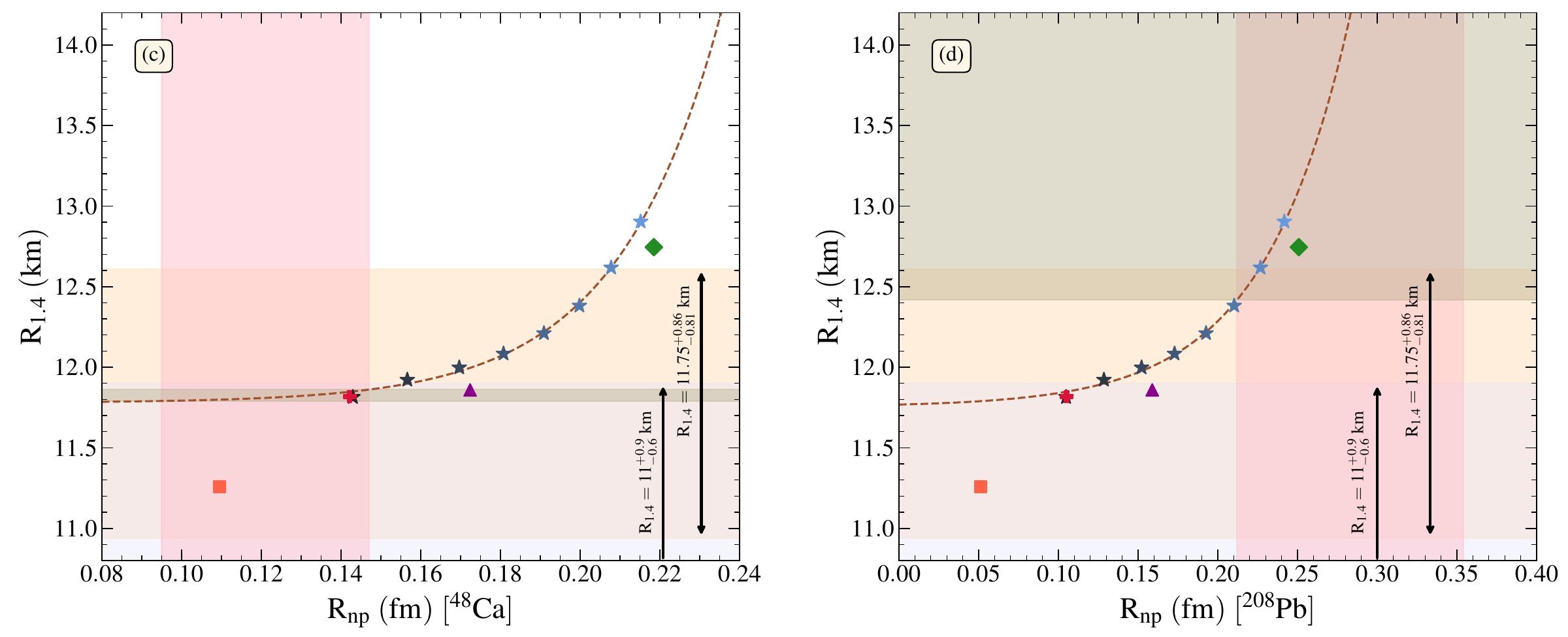}
\caption{Radius at $\rm 1.4~M_{\odot}$ as a function of (a,b) the difference between charge and weak form factors and (c,d) neutron skin thickness for the $\rm ^{48}Ca$ (left panel) and $\rm ^{208}Pb$ (right panel) nuclei. The vertical shaded region represents the experimental constraints extracted from (a,c) CREX and (b,d) PREX-2, while the horizontal shaded regions correspond to constraints extracted from Refs.~\cite{Capano-2020,doi:10.1126/science.abb4317}. The symbols denote different EoSs: DD-PC family of EoSs (stars), DD-PC-CREX (square), DD-PC-PREX (diamond), DD-PC-REX (plus sign), and DD-PCX (triangle). The dashed line represents the fit corresponding to the DD-PC family of EoSs (since $\sigma\ll 0.1$, the standard deviation is not plotted).}
\label{fig:R_ddpc}
\end{figure*}

Fig.~\ref{fig:L_ddpc}(a,b) denotes the dimensionless tidal deformability at $\rm 1.4~M_{\odot}$, $\Lambda_{1.4}$, as a function of the difference between charge and weak form factors $F_{ch}-F_{w}$~\cite{YUKSEL2023137622}, calculated for the DD-PC family of EDFs $(J=29-36~{\rm MeV})$ and for the individual DD-PC interactions constrained in Ref.~\cite{YUKSEL2023137622} by PREX-2 and CREX data. In particular, solid symbols represent the full calculation of the energy per baryon, as detailed in Eq.~\eqref{eq:energy_per_baryon}, for the DD-PC EoSs, incorporating terms up to the fourth order in symmetry energy, $S_{4}(\rho)$. In contrast, the corresponding open symbols denote calculations truncated at the second order term, $S_{2}(\rho)$.
The results reveal an exponential relationship between $\Lambda_{1.4}$ and $F_{ch}-F_{w}$, particularly within the family of DD-PC EoSs. The remaining EoSs generally adhere to this relationship, with DD-PC-CREX and DD-PC-PREX EoSs exhibiting the highest deviations. Moreover, the approaches up to $S_{2}(\rho)$ and  $S_{4}(\rho)$ terms align closely at mid-range values of symmetry energy, $J=31,32~{\rm MeV}$, while at lower and higher values one can observe differences. This behavior is also emphasized in the inset figures containing fit curves for the approach up to the second- and fourth-term in symmetry energy. Although there are no substantial differences observed in the mass-radius plane, tidal deformability is notably influenced by the variations in the EoS as it is related to the low density behavior of the EoS. It is worth noting that the majority of the previous studies have considered calculations only up to the second-order in the symmetry energy, neglecting the comparatively small contribution from the fourth-order term~\cite{universe6080119,TSANG20191,PhysRevC.98.035804,PhysRevD.99.121301}.\\
\indent The results of model calculations, as shown in Fig.~\ref{fig:L_ddpc}(a,b), are compared with constraints on tidal deformability from GW170817 event~\cite{Abbott-2019} (horizontal shaded region), and experimental constraints on weak form factors from CREX and PREX-2 experiments (vertical shaded regions). Given the significant uncertainty in $\Lambda_{1.4}$, all DD-PC interactions considered in the calculations remain consistent with the observational data. Clearly, further observations are necessary to achieve more precise measurements of tidal deformability. Additionally, the constraints from CREX and PREX-2 experiments appear contradictory, where the former one favors the DD-PC-J29 interaction containing the lowest symmetry energy under consideration, while the latter supports the DD-PC-J34,35,36 interactions with high values of symmetry energy, as discussed in Ref.~\cite{YUKSEL2023137622}. Similar discrepancy can also be observed for the two distinct interactions directly constrained by the CREX and PREX-2 data, DD-PC-CREX and DD-PC-PREX, respectively.\\
\begin{table*}
	%\squeezetable
	\caption{Coefficients of expression~\eqref{eq:lambda_f} for $\rm ^{48}Ca$ and $\rm ^{208}Pb$ nuclei for $\mathcal{Q}=\Lambda_{1.4},~{R_{1.4}}$, and $\mathcal{N}= F_{ch}-F_{w},~R_{np}$. In addition, the standard deviation, $\sigma$, and the coefficient of determination $R^{2}$ are also noted. The units of coefficients and $\sigma$ are defined so $\Lambda_{1.4}$ and $F_{ch}-F_{w}$ to be dimensionless, $R_{1.4}$ to be in km, and $R_{np}$ to be in fm.}
    \begin{center}
    \footnotesize
		\begin{tabular}{l l l r r r r r}
            \hline
            \hline
            \multicolumn{5}{c}{\vspace{-0.25cm}} \\
            & Nuclei & Microscopic Properties & $c_{1}$  & $c_{2}$ & $c_{3}$ & $\sigma$ & $R^{2}$\\
            \hline
            \multicolumn{5}{c}{\vspace{-0.25cm}} \\ % Blank row for spacing
            \vspace{0.15cm} \multirow[t]{4}{*}{$\rm \Lambda_{1.4}$} & \multirow[t]{2}{*}{$\rm ^{48}Ca$} & $\rm F_{ch}-F_{w}$ & 94.946 & 32.874 & -- & 14.611 & 0.944\\
            \vspace{0.1cm} & & $\rm R_{np}$ & 97.846 & 7.129 & -- & 14.674 & 0.944\\
            \multicolumn{5}{c}{\vspace{-0.25cm}} \\ % Blank row for spacing
            \vspace{0.15cm} & \multirow[t]{2}{*}{$\rm ^{208}Pb$} & $\rm F_{ch}-F_{w}$ & 200.443 & 23.895 & -- & 14.768 & 0.943\\
			\vspace{0.1cm} & & $\rm R_{np}$ & 185.554 & 3.719 & -- & 14.751 & 0.943\\
            \multicolumn{5}{c}{\vspace{-0.25cm}} \\ % Blank row for spacing
            \vspace{0.15cm} \multirow[t]{4}{*}{$\rm R_{1.4}$} & \multirow[t]{2}{*}{$\rm ^{48}Ca$} & $\rm F_{ch}-F_{w}$ & $3.084\times 10^{-4}$ & 172.174 & 11.772 & 0.019 & 0.997\\
            \vspace{0.1cm} & & $\rm R_{np}$ & $2.931\times 10^{-4}$ & 38.317 & 11.779 & 0.019 & 0.997\\
            \multicolumn{5}{c}{\vspace{-0.25cm}} \\ % Blank row for spacing
            \vspace{0.15cm} & \multirow[t]{2}{*}{$\rm ^{208}Pb$} & $\rm F_{ch}-F_{w}$ & $2.068\times 10^{-2}$ & 116.622 & 11.752 & 0.018 & 0.997\\
			\vspace{0.1cm} & & $\rm R_{np}$ & $1.374\times 10^{-2}$ & 18.282 & 11.754 & 0.018 & 0.997\\
            \hline
		\end{tabular}
   \end{center}
	\label{tab:table1}
\end{table*}
\indent Furthermore, Fig.~\ref{fig:L_ddpc}(c,d) depicts the dimensionless tidal deformability at $1.4~{\rm M_{\odot}}$, $\Lambda_{1.4}$, as a function of the calculated neutron skin thickness, $R_{np}$, defined as the difference between the neutron and proton root-mean-square (RMS) radii $R_{np} = \sqrt{\langle {R_n^2} \rangle} - \sqrt{\langle {R_p^2} \rangle}$, for $\rm ^{48}Ca$ and $\rm ^{208}Pb$, along with the constraints for the $R_{np}$ extracted from the CREX and PREX-2 experiments, respectively. Similar to the trend observed in Fig.~\ref{fig:L_ddpc}(a,b), the data exhibit an exponential relationship, reinforcing the consistency in behavior across these parameters. This is also a consequence of the strong correlation between $F_{ch}-F_{w}$ and $R_{np}$~\cite{YUKSEL2023137622}. Overall, the microscopic properties examined exhibit a consistent trend when correlated with the neutron star $\Lambda_{1.4}$. With the introduction of the neutron skin thickness, the experimental data from the CREX experiment is only consistent with the DD-PC-J29 interaction, while the PREX-2 experiment leads to the acceptance of DD-PC interactions, such as DD-PC-J35 and DD-PC-J36.
The aforementioned results are further supported by the neutron star radius at $1.4~{\rm M_{\odot}}$, $R_{1.4}$, as shown in Fig.~\ref{fig:R_ddpc}, which demonstrates a comparable dependence on the microscopic properties of finite nuclei under investigation, reflecting the trends observed for the dimensionless tidal deformability. Fig.~\ref{fig:R_ddpc} also shows
the experimental constraints extracted from CREX and PREX-2 experiments (vertical shaded regions), and the constraints for the neutron star radius extracted from observation data~\cite{Capano-2020,doi:10.1126/science.abb4317}, indicating that further studies are required to provide tighter constraints.

To quantify the relationships shown in Figs.~\ref{fig:L_ddpc} and~\ref{fig:R_ddpc}, the following expression is employed,
\begin{flalign}
    & \mathcal{Q} = c_{1}\exp({c_{2} {\mathcal{N}}}) + (c_{3}), &
    \label{eq:lambda_f}
\end{flalign}
where $\mathcal{Q}$ stands for $\Lambda_{1.4}$ and $R_{1.4}$, while $\mathcal{N}$ stands for $F_{ch}-F_{w}$ and $R_{np}$. The coefficient $c_{3}$ is only applicable when $\mathcal{Q}=R_{1.4}$. The coefficients $c_{1}$, $c_{2}$, and $c_{3}$, along with the standard deviation of the Eq.~\eqref{eq:lambda_f} and the coefficient of determination $R^{2}$, obtained by the fit to the results based on the DD-PC family of EDFs, are given in Table~\ref{tab:table1}.

\begin{figure}[t!]
\includegraphics[width=\columnwidth]{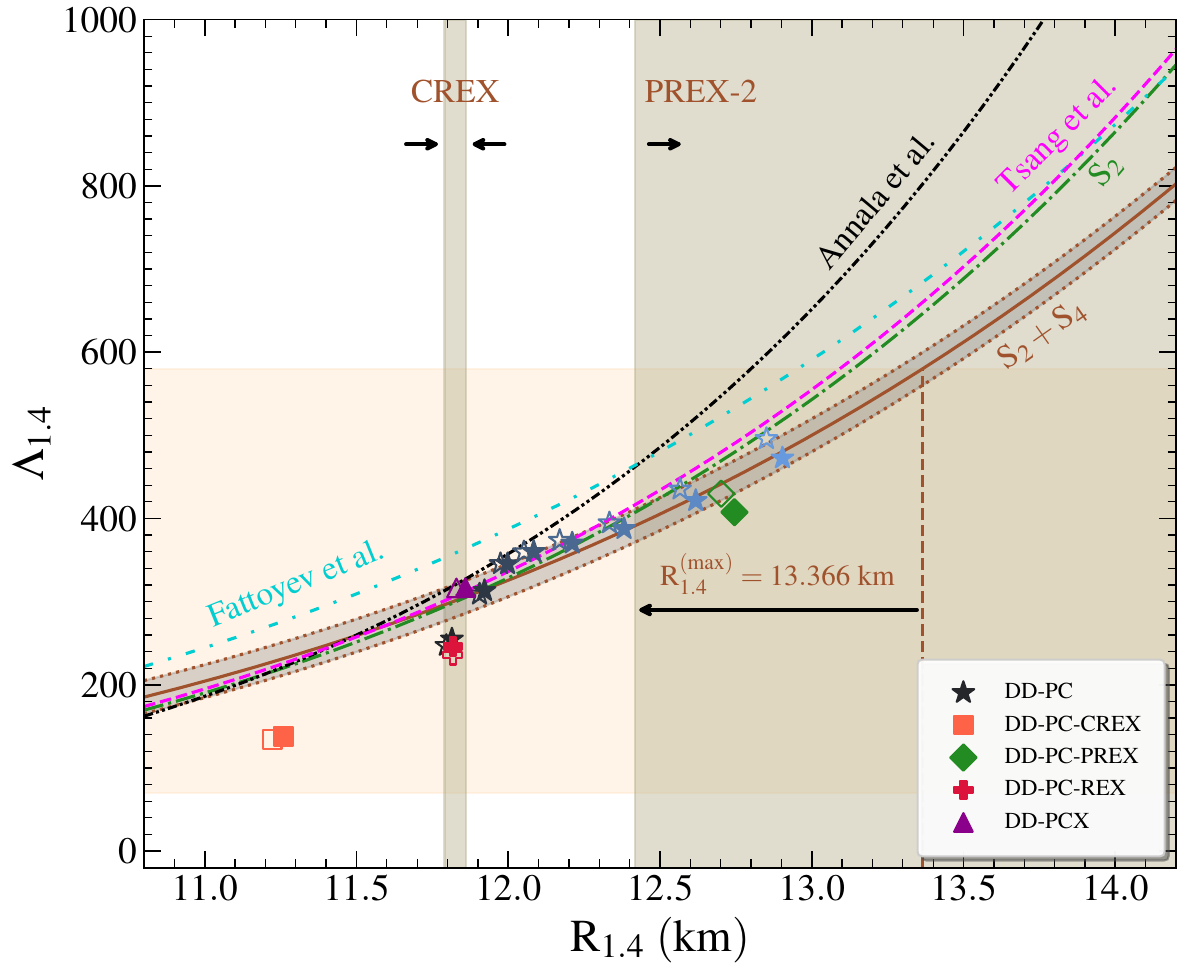}
\caption{Dimensionless tidal deformability as a function of the radius at $\rm 1.4~M_{\odot}$ for the DD-PC EoSs. The horizontal shaded region represents the GW170817 event~\cite{Abbott-2019}, while the vertical shaded regions correspond to the constraints derived in this work. The brown solid line represents the fit for the DD-PC family of EoSs incorporating terms up to the fourth order in the symmetry energy, with the dotted lines denoting the standard deviation. The green dash-dotted line corresponds to the fit for the DD-PC family of EoSs incorporating terms up to the second order in the symmetry energy. Additional comparisons include the black dashed-dot-dotted line from Annala \textit{et al.}~\cite{PhysRevLett.120.172703}, the turquoise dash-dotted line from Fattoyev \textit{et al.}~\cite{PhysRevLett.120.172702}, and the magenta dashed line from Tsang \textit{et al.}~\cite{TSANG20191}}
\label{fig:L_R}
\end{figure}

Utilizing the constraints on $F_{ch}-F_{w}$ from the CREX and PREX-2 experiments, and the extracted neutron skin thickness, along with their established linear dependence with the nuclear symmetry energy at the saturation density and the functional dependence of the quantities of interest introduced in Eq.~\eqref{eq:lambda_f}, we derive boundaries on the neutron star radius at $\rm 1.4~M_{\odot}$ and the symmetry energy. Specifically, the CREX experiment provides a radius range of $11.790\leq R_{1.4} \leq 11.861~{\rm km}$ and a nuclear symmetry energy of $24.306 \leq J \leq 29.037~{\rm MeV}$, while PREX-2 suggests $R_{1.4}\geq 12.417~{\rm km}$ and $J \geq 34.190~{\rm MeV}$. Notably, in the PREX-2 case, due to the well-defined behavior of the density-dependent interaction and its consistency with the GW170817 event at high symmetry energy values ($J=36~{\rm MeV}$ marks the highest value under consideration), we do not impose an upper limit on the radius.\\
\indent To summarize the constraints derived for the nuclear symmetry energy and the properties of neutron stars, Fig.~\ref{fig:L_R} illustrates the dimensionless tidal deformability as a function of radius at $\rm 1.4~M_{\odot}$, along with the extracted constraints from this study. The regions defined by the aforementioned quantities based on CREX and PREX-2 exhibit a contradiction, resulting in a $\rm \sim 0.5~km$ gap between their predictions. In addition, we employ the relation $\Lambda_{1.4} = c_{4}R_{1.4}^{c_{5}}$, where $c_{4}=5.389\times 10^{-4}$ and $c_{5}=5.357$ (correlation coefficient: $r=0.957$, $R^{2}=0.897$) to provide a general behavior of the EoSs based on the relativistic EDFs with the point-coupling interaction. The values of the coefficients are consistent with the strong correlation proposed in Ref.~\cite{PhysRevD.98.063020} and similar studies in Refs.~\cite{universe6080119,TSANG20191,PhysRevC.98.035804,PhysRevLett.120.172703,PhysRevLett.120.172702}. By applying the upper limit for the dimensionless tidal deformability at $\rm 1.4~M_{\odot}$, the radius is limited to $R_{1.4}\leq 13.366~{\rm km}$, which is in agreement with previous studies~\cite{TSANG20191,PhysRevC.98.035804,PhysRevLett.120.172703,PhysRevLett.120.172702}. Furthermore, as this upper limit lies within the region defined by PREX-2 experiment, we adopt this value as the upper limit to impose an upper limit for the nuclear symmetry energy, which is obtained in the range $34.190 \leq J \leq 36.658~{\rm MeV}$. The results for the neutron star radius at $\rm 1.4~M_{\odot}$ and the nuclear symmetry energy at the saturation density are provided in Table~\ref{tab:table2}. We note that while the experimental uncertainties derived from CREX and PREX-2 are comparable, the corresponding constraints on radius and symmetry energy exhibit different behaviors. Specifically, the CREX band for the neutron star radius is narrower than the PREX-2 band. This discrepancy arises from CREX favoring lower symmetry energy values, which correspond to the lower part of the exponential curve where the slope is more gradual. Conversely, PREX-2 favors higher symmetry energy values, corresponding to the upper part of the exponential curve, characterized by a steeper slope.

\begin{table}
	\caption{Neutron star radius at $\rm 1.4~M_{\odot}$ and nuclear symmetry energy at the saturation density obtained using constraints from CREX and PREX-2 experiments.}
    \begin{center}
    \footnotesize
		\begin{tabular}{l c c c c}
            \hline
            \hline
            \multicolumn{5}{c}{\vspace{-0.25cm}} \\
            & \multicolumn{2}{c}{$\rm R_{1.4}~(km)$} & \multicolumn{2}{c}{$\rm J~(MeV)$} \\
            & min & max & min & max \\
            \hline
            \multicolumn{5}{c}{\vspace{-0.25cm}} \\ % Blank row for spacing
            \vspace{0.15cm} CREX & $11.790_{-0.019}^{+0.019}$ & $11.861_{-0.019}^{+0.019}$ & $24.306_{-0.233}^{+0.233}$ & $29.037_{-0.253}^{+0.253}$ \\
            \vspace{0.1cm} PREX-2 & $12.417_{-0.018}^{+0.018}$ & $13.366_{-0.087}^{+0.084}$ & $34.190_{-0.186}^{+0.186}$ & $36.658_{-0.186}^{+0.186}$ \\
            \hline
		\end{tabular}
   \end{center}
	\label{tab:table2}
\end{table}

For comparison, Fig.~\ref{fig:L_R} also shows functional dependencies for $\Lambda_{1.4} - R_{1.4}$ from other studies~\cite{universe6080119,TSANG20191,PhysRevLett.120.172703,PhysRevLett.120.172702}. The present calculation including up to $S_{2}(\rho)$ term appears very similar to previous studies, which also assume expansion up to the second order in the symmetry energy~\cite{universe6080119,TSANG20191}. The complete 
calculation up to fourth order in the symmetry energy, $S_{4}(\rho)$, is rather different than presented results from previous studies. Generally, the inclusion of $S_{4}(\rho)$ term in the symmetry energy results in larger radii compared to the $S_{2}(\rho)$ approach, with this effect becomes more pronounced at higher symmetry energy values. Furthermore, the corresponding tidal deformability exhibits a distinct trend based on the symmetry energy: (a) for low symmetry energy ($J\leq 30~{\rm MeV}$) the inclusion of $S_{4}(\rho)$ leads to higher values of tidal deformability, (b) for intermediate symmetry energy ($30 < J\leq 32~{\rm MeV}$) the inclusion of $S_{4}(\rho)$  yields tidal deformability values similar to those from the $S_{2}(\rho)$ approach, and (c) for higher symmetry energy ($J> 32~{\rm MeV}$) the inclusion of $S_{4}(\rho)$ leads to lower values of tidal deformability. In addition, from the derived fits, two key trends emerge regarding the interplay between radius and tidal deformability. For a fixed tidal deformability, the $S_{4}(\rho)$ approach results in lower radii for $J\lesssim 30~{\rm MeV}$, whereas for $J\gtrsim 30~{\rm MeV}$, it yields higher radii. Conversely, for a fixed radius, the $S_{4}(\rho)$ approach leads to higher tidal deformability values for $J\lesssim 30~{\rm MeV}$ and lower values for $J\gtrsim 30~{\rm MeV}$.

To provide a neutron star description that is more directly consistent with findings from both the CREX and PREX-2 experiments, we also present results from three additional EoSs: DD-PC-CREX, DD-PC-PREX, and DD-PC-REX functionals~\cite{YUKSEL2023137622}. Although the first two EoSs successfully satisfy both microscopic and macroscopic constraints, they deviate from the DD-PC family of systematically obtained EoSs. In contrast, DD-PC-REX better aligns with the characteristics of the DD-PC family while simultaneously satisfying both the CREX and PREX-2 constraints. Notably, the DD-PC-REX EoS exhibits relatively low symmetry energy, closely approaching the upper edge of the CREX experimental range.

\section{Concluding remarks}
In conclusion, the connection between nuclear matter properties and neutron star observables facilitates the relationship between terrestrial experiments on finite nuclei and astrophysical observations of neutron stars. 
Specifically, this work utilizes the set of DD-PC interactions within the relativistic EDF framework, spanning a range of symmetry energy values $J$, along with the charge and weak form factors obtained from the CREX and PREX-2 experiments and the tidal deformability extracted from neutron star merger observation data. In this work, the energy density in the EoSs is extended to include the second- and fourth-order terms of the symmetry energy. These two approaches result in differences not only in the microscopic properties of neutron stars, such as the crust-core transition density and proton fraction~\cite{PhysRevC.85.024302,PhysRevC.80.014322}, but also in macroscopic characteristics, including neutron star radius and tidal deformability. The increase in neutron star radius and the distinct behavior of tidal deformability, considering a neutron star of $1.4~{\rm M_{\odot}}$, highlight the necessity of including higher order terms in the symmetry energy. Although the impact of these terms is comparatively smaller than the second-order term, their inclusion provides a more accurate description of dense matter properties and contributes to improving the agreement between theoretical predictions and astrophysical observations.

The DD-PC framework and the resulting EoSs are consistent with prior studies and reaffirm the discrepancy between the constraints provided by the CREX and PREX-2 experiments. The CREX suggests a softer EoS at saturation density, whereas PREX-2 favors a stiffer EoS. Furthermore, the DD-PC framework is employed to derive constraints on $R_{1.4}$ and $J$ by utilizing the calculated functional dependencies between the microscopic properties of finite nuclei and the macroscopic characteristics of neutron stars. Finally, three additional EoSs are discussed, which are established directly from the measured weak form factors. Notably, the DD-PC-REX EoS aligns with the general trend observed in the DD-PC family and provides well-defined values for $R_{1.4}$ and $J$ within the region defined by CREX.

Further experiments, such as the upcoming parity-violating electron scattering MREX, have the potential to provide more precise constraints on the weak form factor, and, as a consequence, smaller uncertainties in extracted neutron skin thickness and symmetry energy. The MREX experiment could also open perspectives to reduce the symmetry energy and radius gap between the CREX and PREX-2 results. These advancements will deepen our understanding of neutron star properties and improve the accuracy of existing models by providing more precise data on the isovector components of nuclear matter.

\section*{Acknowledgments}
This work is supported by the Croatian Science Foundation under the project number HRZZ-MOBDOL-12-2023-6026 and under the project Relativistic Nuclear Many-Body Theory in the Multimessenger Observation Era (HRZZ-IP-2022-10-7773). E.Y. acknowledges support from the UK STFC under award no. ST/Y000358/1.
%==================================================================================================================================================
\bibliographystyle{elsarticle-num}
\bibliography{bibliography}
%==================================================================================================================================================

\end{document}